\documentclass[twocolumn,showpacs,preprintnumbers,amsmath,amssymb,superscriptaddress]{revtex4}
\usepackage{epsfig,bm,color}
\usepackage{ulem}
\usepackage{amsmath}
\begin{document}
\title{Electric dipole response of $^6$He: Halo-neutron and core excitations}
\author{D. Mikami\footnote{Present address: Hitachi Solutions East Japan, Ltd.,
Sendai 980-0014, Japan}}
\affiliation{Department of Physics, Hokkaido University, Sapporo 060-0810, Japan}
\author{W. Horiuchi}
\affiliation{Department of Physics, Hokkaido University, Sapporo 060-0810, Japan}
\author{Y. Suzuki}
\affiliation{Department of Physics, Niigata University, Niigata 950-2181, Japan}
\affiliation{RIKEN Nishina Center, Wako 351-0198, Japan}

\begin{abstract}
Electric dipole ($E1$) response of $^{6}$He is studied with 
a fully microscopic six-body calculation.
The wave functions for the ground and excited states are expressed 
as a superposition of explicitly correlated Gaussians (CG).
Final state interactions of three-body decay channels
are explicitly taken into account. 
The ground state properties and 
the low-energy $E1$ strength are obtained  consistently with observations.
Two  main peaks as well as several small peaks are found in the $E1$ strength function. 
The peak at the high-energy region indicates a typical macroscopic picture of the giant 
dipole resonance,  the out-of-phase proton-neutron motion.
The  transition densities of the lower-lying peaks exhibit in-phase proton-neutron motion in the internal region, out-of-phase motion near the surface 
region, and spatially extended neutron oscillation, indicating a soft-dipole mode (SDM) and its vibrationally excited mode. 
\end{abstract}

\maketitle

\section{Introduction}

Exploring new phenomena in unstable nuclei has become possible 
due to intense radioactive beams produced by new facilities.
A neutron halo is one of the most attractive phenomena
found in some unstable nuclei near the neutron dripline,
and has  attracted  much attention
since the discovery of the large matter radius of $^{11}$Li~\cite{tanihata85}.
Typical two neutron halo nuclei
are, e.g., $^6$He~\cite{tanihata85b}, $^{11}$Li, $^{14}$Be~\cite{tanihata88}, 
and the recently observed $^{22}$C~\cite{tanaka10}.
A common feature of these nuclei 
is a small two-neutron separation energy ($S_{2n}$)
that leads to a large matter radius.

It is known that such weakly bound systems exhibit
large $E1$ strength at the low-energy region as often studied 
through  Coulomb breakup reactions~\cite{aumann99,nakamura06,kobayashi12}.
As a unique phenomenon in the neutron rich nuclei, the possibility  
of the SDM has for a long time been discussed 
as a vibration of the halo neutrons against the core~\cite{hansen87,ikeda88,suzuki90a}, which  
is a variant of the macroscopic picture 
of the giant dipole resonance (GDR) such as   
Goldhaber-Teller~\cite{goldhaber48}
and Steinwedel-Jensen~\cite{steinwedel50} models.

These days low-lying $E1$ strength in medium- and heavy-mass nuclei
has been studied extensively in its relationship 
to neutron-skin thickness and neutron matter 
in a neutron star~\cite{chen10, inakura13}.
The low-lying strength is observed in neutron-proton unbalanced nuclei
and it is often called  a pygmy dipole resonance (PDR) because its strength is 
much smaller than that of the GDR.
However, its excitation mechanism is still controversial concerning 
whether the mode is SDM, collective or single-particle excitation.

In this paper, we study the $E1$ response of $^{6}$He up to the GDR region, 
focusing on the possibility of the SDM as well as other $E1$ excitation modes.
The low-lying $E1$ strength of $^6$He was observed by  
the Coulomb breakup experiment of $^6$He~\cite{aumann99}, which found 
a broad peak in the low-energy region just few MeV 
above the $\alpha+n+n$ threshold.  
An indication of the low-lying peak is also reported in a 
$^6$Li($^7$Li,$^7$Be)$^6$He charge-exchange reaction~\cite{nakayama00}.
Theoretically the $E1$ response of $^6$He  
is often studied with a macroscopic $\alpha+n+n$ three-body 
model~\cite{suzuki91,danilin97,myo01,baye09,hagino09,pinilla11}. However, some  
theoretical uncertainty exists when the $\alpha$ particle is treated as a point particle.
Even if the $\alpha+n+n$ three-body problem is solved accurately, two phase-shift equivalent 
$\alpha-n$ potentials give different $E1$ strength as shown in Refs.~\cite{myo01,baye09,pinilla11}.
To avoid such uncertainties, 
we study the $^{6}$He nucleus in a fully microscopic six-nucleon calculation.
The six-body model has another important advantage that the distortion of 
the $\alpha$ core is naturally taken into account.
The distortion or core polarization effect is known to play
an important role in binding the halo neutrons of $^6$He~\cite{arai99}. 
Bacca {\it et al.}\, presented six-body calculations for $^{6}$He
with the effective interaction hyperspherical harmonics (EIHH) 
combined with a Lorentz integral transform~\cite{bacca02, bacca04}. 
They obtain two peaks for $^{6}$He and speculate 
the existence of the SDM at the low-lying peak as well as 
the GDR at the higher peak.  

The paper is organized as follows.
In Sec.~\ref{model.sec}, our six-body model is formulated.
The Hamiltonian, basis functions, and model space adopted in the present work
are explained here. We perform a variational calculation to obtain the 
ground state wave function. Those configurations that are accessible by 
the $E1$ operator are carefully prepared to take account of all the $E1$ 
strength of $^6$He. 
Results and discussions are presented in Sec.~\ref{results.sec}.
First, we show the ground state properties of $^6$He
and discuss its structure in Sec.~\ref{ground.sec}.
Next we present the $E1$ strength  
and  analyze important configurations for describing the $E1$ excitation
in Sec.~\ref{excitation.sec}. A comparison with experimental data 
is made in Sec.~\ref{photo.sec}.  
The low-lying $E1$ excitation mode as well as 
the one in the GDR region are 
discussed in detail in Sec.~\ref{softmode.sec}.
Sec.~\ref{three-body.sec} discusses the extent to which the $\alpha+n+n$ three-body picture is validated 
for $^{6}$He. Result on  compressional $E1$ strength is shown in Sec.~\ref{comp.E1}. 
Conclusions are drawn in Sec.~\ref{conclusions.sec}.

\section{Method}

\label{model.sec}

\subsection{Hamiltonian}

The Hamiltonian of an $N$-particle system is specified by
a kinetic energy and an effective two-body interaction between nucleons: 
\begin{align}
H=\sum_{i=1}^{N} T_i-T_\text{cm}+\sum_{i<j}v_{ij}.
\end{align}
The proton-neutron mass difference is ignored in the kinetic energy.
The center-of-mass (c.m.) motion of the total system, $T_{\rm cm}$, is excluded,
and no spurious c.m. motion appears in the calculation.
As the two-body interaction, 
we employ the central Minnesota (MN) potential~\cite{MN} that  fairly well 
reproduces the binding energies of $N=2\sim 6$ systems~\cite{varga95}. 
No three-body force is included. 
A spin-orbit interaction is often employed, in addition to the central MN force,
to reproduce the splitting of $p_{3/2}$ and $p_{1/2}$ phase shifts 
of $^{4}$He$+n$ system~\cite{reichstein70}. That interaction was applied to 
describe the neutron halo structure of $^{6,8}$He in a microscopic 
three-body and five-body model~\cite{varga94a}. 
We use, however, only the central MN force because the effective spin-orbit force 
gives ill behavior as shown in a four-body calculation for $^{4}$He~\cite{horiuchi13b}.  
Instead we adjust the strength of odd partial waves 
 by changing  one free parameter, $u$, of the MN potential.
The $u$ parameter does not affect the binding energies
of $N<5$ systems that are mainly composed of $S$-state. 
A choice of $u$ will be discussed later. The Coulomb potential is included. 
The nucleon mass $m_N$ and the charge constant $e$ used in what follows  
are $\hbar^2/m_N=41.47$ MeV\,fm$^2$ and $e^2=1.440$ MeV\,fm.

\subsection{Basis functions for bound states}

In this work, 
we solve a many-body Schr\"{o}dinger equation using a variational method.
A bound-state solution of the $N$-nucleon system with spin-parity $J^\pi$  
is expressed in terms of a linear combination of the $LS$ coupled 
basis functions 
\begin{align}
\Phi_{(LS)JM_JM_T}^{(N)\pi}=\mathcal{A}\left\{\left[\phi_L^{(N)\pi}
\chi_S^{(N)}\right]_{JM_J}\eta_{M_T}\right\},
\label{lscoupled}
\end{align}
where $\mathcal{A}$ is the antisymmetrizer, and the 
square bracket denotes the tensor product of angular momentum coupling. 
The spin function $\chi_S$ is given  
in a successive coupling as 
\begin{align}
&\chi_{S_{12},S_{123},\dots, S M_S}^{(N)}\notag\\
&=[\dots[[\chi_\frac{1}{2}(1) \chi_\frac{1}{2}(2)]_{S_{12}}
 \chi_\frac{1}{2}(3)]_{S_{123}}\dots]_{SM_S}.
\label{spin.fn}
\end{align}
Note that the above spin function forms 
a complete set provided all
possible intermediate spins ($S_{12}, S_{123}$, $\dots$) 
are included for a given $S$. 
The isospin function $\eta_{M_T}$ is expressed with a product of 
single-particle isospin functions. 
The function $\eta_{M_T}$ 
can also be expressed by a linear combination of Eq.~(\ref{spin.fn}),
e.g., with total isospin $T=0$, 1, 2 for $^4$He  
and $T=1$, 2, 3 for $^6$He.

A choice of the variational trial functions is essential to determine
the accuracy of the calculation. We employ
the CG basis~\cite{boys60,singer60},
which is flexible to treat few-body dynamics, e.g., to 
describe a tail in the asymptotic region as well as clustering~\cite{horiuchi08, horiuchi14}. 
Also see a recent review~\cite{mitroy13} 
for various powerful applications of the CG.
Denoting the nucleon coordinate by $\bm r_i$, we use a set of the Jacobi coordinates, 
$\bm{x}_i=\bm{r}_{i+1}-\sum_{j=1}^{i}\bm{r}_{j}/i$ ($i=1,\dots,N-1$) but
other sets of relative coordinates may be used as well.  
We introduce a short-hand notation  $\bm{x}$ that is an $(N-1)$-dimensional column vector or 
an $(N-1)\times 1$ matrix whose $i$th element is the 3-dimensional vector $\bm{x}_i$. 
The spatial part $\phi_{L}^{(N)\pi}$ of Eq.~(\ref{lscoupled}) 
generally takes the form~\cite{varga95,svm}
\begin{align}
&F_{LM_L}(v, A, \bm{x})=\exp(-\textstyle{\frac{1}{2}}\tilde{\bm{x}}A\bm{x})
\mathcal{Y}_{LM_L}(\tilde{v}\bm{x})
\label{GVR.eq}
\end{align}
with a solid harmonic
\begin{align}
\mathcal{Y}_{\ell m}(\bm r)=r^{\ell}Y_{\ell m}(\hat{\bm r}),
\end{align}
where $A$ is an $(N-1)\times (N-1)$ positive-definite, symmetric matrix
and $\tilde{\bm{x}}A\bm{x}$ stands for
$\sum_{i,j=1}^{N-1}A_{ij}\bm{x}_i\cdot\bm{x}_j$.
The tilde indicates the transpose of a matrix.
The parameter $v$ is $(N-1)$-dimensional column vector that 
defines a global vector (GV), $\tilde{v}\bm x(=\sum_{i=1}^{N-1}{v}_i\bm x_i)$, 
which is responsible for describing the angular motion of the system.
For $^6$He with the central MN potential,
we only need to consider the lowest $L$ because no channel coupling occurs 
between states with different $L$:
$L=0$ for the ground states and $L=1$ for the $E1$ excited states.

The CG-GV basis~(\ref{GVR.eq}) explicitly describes correlated motion 
among the particles through the off-diagonal elements of $A$ 
and the rotational motion of the system is conveniently described by the GV. 
Both bound and excited states are expressed by the same functional form. 
For the bound states, the variational parameters are determined by the 
stochastic variational method (SVM)~\cite{varga94b,varga95,svm}. 
The wave functions for the excited states are constructed
on the basis of single particle (sp) or cluster excitations as 
explained in Sec.~\ref{wavefn.sec}. 
Most noticeable among several advantages of the CG-GV basis functions is that 
the functional form of Eq.~(\ref{GVR.eq}) 
remains unchanged under an arbitrary linear 
transformation of the coordinate $\bm x$. 
Another advantage is that 
the matrix elements for most operators 
can be evaluated analytically, 
which allows us to obtain the matrix elements 
accurately with a low computational cost.
Useful formulas for evaluating matrix elements with the CG-GV basis 
are collected in Appendices of Refs.~\cite{suzuki08,aoyama12}.

\subsection{Basis functions for electric dipole excitation}

\label{wavefn.sec}

\begin{figure}
\begin{center}
\epsfig{file=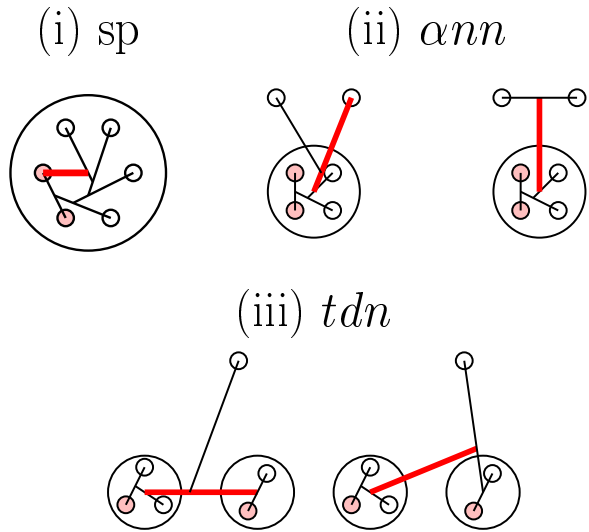,scale=0.75}
\caption{(Color online). Schematic diagrams of three patterns 
for the $E1$ excitations of $^6$He.
Small circles and shaded ones denote neutron and proton, respectively. 
Thick solid lines denote the coordinates 
on which the $E1$ operator acts.}
\label{e1ext.fig}
\end{center}
\end{figure}

We construct basis functions for the final states
with $J^\pi=1^-$ that are excited by the $E1$ operator 
\begin{align}
\mathcal{M}_{1\mu}=e\sum_{i\in p} (\bm{r}_i-\bm{x}_6)_{\mu}=\sqrt{\frac{4\pi}{3}}e\sum_{i\in p} {\cal Y}_{1\mu}(\bm{r}_i-\bm{x}_6),
\label{E1.eq}
\end{align}
where the summation runs over protons, and
$\bm{x}_6$ is the c.m. coordinate of $^6$He. We apply the same prescription as that of 
Refs.~\cite{horiuchi12a, horiuchi13a} adopted for the study of $E1$ and spin-dipole strength in 
$N=4$ system to the six-body calculation of $E1$ strength. In that study 
four-body continuum-discretized states are constructed by taking into account 
both sum rules and final state interactions, and thereby the model 
reproduces experimental data satisfactorily and 
in addition leads to some predictions.
The  configurations that we include here are
(i) the sp excitation  built with 
$\mathcal{Y}_{1\mu}(\bm{r}_1-\bm{x}_6)$,
(ii) the $\alpha+n+n$ three-body decay channel,
and (iii) the $t+d+n$ three-body decay channel. 
Figure~\ref{e1ext.fig} illustrates schematic diagrams of the above three configurations
for the $E1$ excitation of $^6$He.
Because of a Borromean nature of $^6$He, an 
explicit inclusion of the three-body final states is important.
The basis (i) is  vital to satisfy the sum rule,
and the bases (ii) and (iii) take care of final state interactions 
in the three-body decay asymptotics. The detail of 
each configuration is given below. 
It should be noted, however, that the three classes of configurations
are not orthogonal to each other but have considerable overlap among others.

\subsubsection{Single-particle excitation}

We define the sp basis by  
\begin{align}
\Psi_f^{\text{sp}}=\mathcal{A}\left[\Phi_{0,i}^{(6)}\mathcal{Y}_1({\bm{r}}_1-\bm{x}_6)\right]_{1\mu},
\end{align}
where $\Phi_{0,i}^{(6)}$ is the $i$th component of  
the ground state wave function of $^{6}$He. 
The basis state describes a sp excitation 
from the ground state by the $E1$ operator and plays an important role in accounting 
for the $E1$ sum rule. If $\Phi_{0,i}^{(6)}$ is replaced with the ground state 
wave function of $^6$He, $\Psi_f^{\text{sp}}$ reduces to the coherent state that   
exhausts all the $E1$ strength reached from the ground state.  
Since all the components are included independently in the present calculation, 
the set of the basis functions takes into account the effect of the pseudo excited $0^+$ states of $^{6}$He. 

\subsubsection{$\alpha+n+n$ three-body decay channel}

The $\alpha+n+n$ channel is the lowest threshold of $^{6}$He 
and is expected to be important to describe the SDM.
We explicitly include the $\alpha+n+n$ configurations in the form 
\begin{align}
\Psi_f^{\alpha n n}=\mathcal{A}
\left[\Phi_{0,i}^{(4)}\exp\left(-\tfrac{1}{2}\tilde{\bm{y}}B\bm{y}\right)
\left[\mathcal{Y}_1(\tilde{w}\bm{y})\chi^{(2)}_{S_{56}}\right]_1\right]_{1\mu},
\label{type(ii)}
\end{align}
where $\Phi_{0,i}^{(4)}$ is the $i$th basis state of the ground state wave function
of $^{4}$He, and $\tilde{\bm{y}}=(\bm {y}_1\, \bm{y}_2)$ 
where the coordinates $\bm y_1$ and 
$\bm y_2$  specify the three-body character of $\alpha+n+n$ and so-called Y- 
and T-type coordinates are both employed. 
Further detail is given in the next paragraph. 
See Fig.~\ref{e1ext.fig} (ii).
A $2\times 2$ positive-definite, symmetric matrix $B$ that characterizes
a spatial configuration of the two valence neutrons  is chosen
to cover physically important region for describing
weakly bound two neutrons. A choice of $w$ is also explained below. 
The spin of the two neutrons, $S_{56}$, 
can be 0 and 1, and both are included independently in the basis set.
The isospin functions are not written explicitly in Eq.~(\ref{type(ii)}) for the sake of simplicity.
In order to reduce a computational cost,
we prepare a truncated $^4$He wave function with 15 basis states, 
which leads to an energy loss of 0.3\,MeV.
We also calculate the $E1$ strength with 20 and 25 basis states and  confirm that 
no qualitative difference is obtained in the $E1$ strength below 20 MeV excitation energy 
where the $\alpha$ core remains in its ground state.
Since the basis state $\Phi_{0,i}^{(4)}$ is included independently, the two neutrons are allowed 
to move against the $\alpha$ core in its pseudo excited state. 

Both Y- and T-type coordinates are defined by 
\begin{align}
\begin{split}
\bm{y}^{\rm (Y)}:\ 
&\bm{y}_1^{\rm (Y)}=\bm{r}_5-\bm{x}_{\rm cm}^{(4)},\ \ 
\bm{y}_2^{\rm (Y)}=\bm{r}_6-\frac{\bm{r}_5+4\bm{x}_{\rm cm}^{(4)}}{5},\\
\bm{y}^{\rm (T)}:\
&\bm{y}_1^{\rm (T)}=\bm{r}_6-\bm{r}_5,\ \ 
\bm{y}_2^{\rm (T)}=\frac{\bm{r}_5+\bm{r}_6}{2}-\bm{x}_{\rm cm}^{(4)},
\end{split}
\end{align}
where $\bm{x}_{\rm cm}^{(N)}$ denotes the c.m. coordinate of an $N$-particle subsystem. 
Since the main neutron configuration in the ground state of $^6$He is $(0s)^2(0p)^2$,
the $E1$ operator acting on the halo neutron changes one of the $P$ orbits to an extended $S$ orbit 
in the continuum. In the case of Y-type that describes a valence-neutron excitation 
the $S$-wave is assigned to $\bm{y}^{\rm (Y)}_1$ and the remaining $P$-wave to $\bm{y}^{\rm (Y)}_2$. 
In the case of T-type that is very important to describe the SDM-like excitation the $S$-wave is 
assigned to $\bm{y}^{\rm (T)}_1$ while the $P$-wave to $\bm{y}^{\rm (T)}_2$. The above assignment of the 
partial waves can be made possible by choosing $\tilde{w}=(0\, 1)$ and $B$ to be diagonal. 

In practical calculations, first we generate the diagonal matrix elements of $B$,
and then transform the coordinates $\bm y^{(\rm Y)}$ or $\bm y^{(\rm T)}$ together with 
three coordinates used to describe $\Phi_{0,i}^{(4)}$ to the $\bm x$ coordinate. This 
transformation is carried out by an appropriate 
matrix $T$. Substitution of $\bm y=T\bm x$ into 
Eq.~(\ref{type(ii)}) reduces the $\alpha+n+n$ configurations to the standard CG-GV basis function 
of Eq.~(\ref{lscoupled}). The diagonal elements, $B_{11}$ and $B_{22}$, 
are taken by a geometrical progression
with different Gaussian falloff parameters 
ranging from 0.1 to 22 fm: More explicitly, they are
chosen as $B_{11}=(0.13\times 1.35^{(n-1)})^{-2}\ (n=1,\dots 18)$ and
$B_{22}=(0.2\times 1.4^{(m-1)})^{-2}\ (m=1,\dots 15)$ in both Y- and T-types.

\subsubsection{$t+d+n$ three-body decay channel}

We here treat the $t+d+n$ three-body decay channel. This channel 
is important to describe the $E1$ strength  
especially in the GDR region because it involves the excitation of the $\alpha$ core. 
The channel is the third lowest threshold of $^6$He and makes it possible to 
describe those configurations in which  
two protons, excited by the $E1$ operator, are apart from each other. 
With the coordinate sets appropriate for describing the three-body system 
\begin{align}
\begin{split}
\bm{z}^{\rm (Y)}:\ 
&\bm{z}^{\rm (Y)}_1=\bm{x}_{\rm cm}^{(2)}-\bm{x}_{\rm cm}^{(3)},\ \
\bm{z}^{\rm (Y)}_2=\bm{r}_6-\frac{2\bm{x}_{\rm cm}^{(2)}+3\bm{x}_{\rm cm}^{(3)}}{5},\\
\bm{z}^{\rm (T)}:\
&\bm{z}^{\rm (T)}_1=\frac{\bm{r}_6+2\bm{x}_{\rm cm}^{(2)}}{3}-\bm{x}_{\rm cm}^{(3)},\ \
\bm{z}^{\rm (T)}_2=\bm{r}_6-\bm{x}_{\rm cm}^{(2)},
\end{split}
\end{align}
the basis functions are expressed as 
\begin{align}
\Psi_f^{tdn}
&=\mathcal{A} \left[\left[ \Phi_{1/2,i}^{(3)}
\Phi_{1,j}^{(2)}\right]_{J_{12}}\right.
\notag\\
&\times\left. \exp\left(-\tfrac{1}{2}\tilde{\bm{z}}B\bm{z}\right)
\left[
\mathcal{Y}_1(\tilde{w}\bm{z})\Phi_{1/2}^{(1)}\right]_{J_{34}}\right]_{1\mu},
\end{align}
where $\tilde{\bm{z}}=(\bm{z}_1 \, \bm{z}_2)$, 
and $\Phi_{1/2,i}^{(3)}$ and $\Phi_{1,j}^{(2)}$
are $i$th and $j$th basis functions of $^3$H and $^2$H, 
respectively. They are approximated by 7  for $^3$H
and 3 basis states for $^2$H.
The  $\Phi_{1/2}^{(1)}$ represents the single
neutron spin and isospin function. 
The intermediate spins, $J_{12}$ and $J_{34}$, take 1/2 and 3/2, 
and they are included independently. 
The partial wave of the first relative coordinate in each coordinate set
is chosen to be a $P$-wave that is excited by the $E1$ operator.
The basis states take  into account  the asymptotics of 
the three-body decay due to the $E1$ excitation 
as well as a coupling with the pseudo states of $^3$H and $^2$H.
Similarly to the construction of $\alpha+n+n$ basis states, 
we set $\tilde{w}=(1\, 0)$ and choose the diagonal matrix element of $B$ as
$B_{11}=(0.7\times 2.0^{(n-1)})^{-2}\ (n=1,\dots,5)$
and $B_{22}=(0.7\times 2.0^{(m-1)})^{-2}\ (m=1,\dots,5)$. 

Note that the $t+t$ two-body decay channel, the second lowest threshold of $^6$He,  
does not contribute to the $E1$ strength in the present model. The 
$t+t$ channel with the relative $P$-wave motion has to have $S=1$ and $T=1$. Since the ground state of 
$^6$He is described with the $S=0$ configurations and the $E1$ operator 
does not change the spin, any $t+t$ configurations with $S=1$ can not be excited by $E1$. 
If a realistic nuclear force is 
used, both $S=0$ and $S=1$ configurations couple and the $t+t$ channel gives some 
contribution to the $E1$ strength. 

The number of basis functions in each diagram    
is 600 for sp, 4050$\times 2$ for $\alpha+n+n$, and 2625$\times 2$
for $t+d+n$.
The calculation is performed not only in each basis set
but also in a `Full' model space that combines all of them. 
The total number of basis functions in the Full calculation is 13950. Since 
all $E1$ strength of $^6$He exists in the continuum above the $\alpha+n+n$ threshold, our calculation  
is practically an approximation with the continuum discretization. Because of the extensive basis states, however, the number of discretized 
states is so dense that on average 10 states appear per 1\, MeV below the excitation energy of 50\,MeV. 

\section{Results and discussions}

\label{results.sec}

\subsection{Ground state properties}
\label{ground.sec}

\begin{figure}[t]
\begin{center}
\epsfig{file=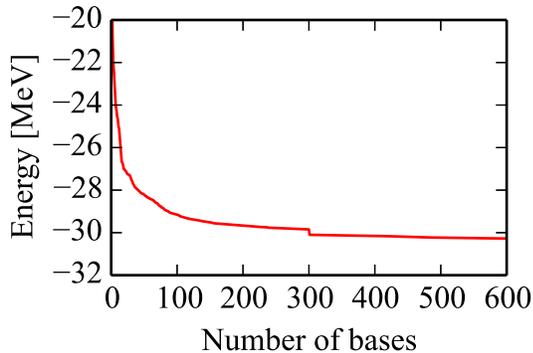,scale=1.}
\caption{(Color online). Energy convergence of $^{6}$He as a function 
of basis dimension. The parameter $u$ of the MN potential is set to be unity.}
\label{conv.fig}
\end{center}
\end{figure}

The CG-GV basis with the SVM optimization gives an efficient description
of the ground state of $^{6}$He.
Figure~\ref{conv.fig} displays the convergent curve of 
the ground state energy of $^6$He calculated with $u=1.0$ of the MN potential.
We increase the number of basis functions competitively according to 
the algorithm of the SVM~\cite{varga95, svm}.
A sudden decrease of the energy at the 300th basis state 
is due to the SVM refinement procedure in which the basis dimension is fixed but each basis function is  compared with the best one among randomly generated candidates. 
The converged energy is obtained only with 600 basis states,
which is surprisingly small if one recalls that 
each basis function has 15 variational parameters for the orbital part (\ref{GVR.eq}). No decrease of the energy is obtained 
when we make the refinement of the 600 basis states.
The calculated ground state energy of $^{6}$He with $u=1.0$ is consistent with
that obtained by the calculation of 600 dimension in Ref.~\cite{varga95}. 
In fact it is   
$0.25$ MeV lower than that because of the improvement of the asymptotics of the wave function. The root mean square (rms) matter radius, $r_m$, is found to be larger by 0.08\,fm than that of Ref.~\cite{varga95}.

\begin{table}[ht]
\begin{center}
\caption{Energy, two-neutron separation energy,
and  rms radii of of $^4$He and $^6$He. See text for details.
Energy and length are given in units of MeV and fm, respectively.
Experimental data are taken from Ref.~\cite{audi03}.
}
\begin{tabular}{ccccccccc}
\hline\hline
      &&&\multicolumn{5}{c}{$u$}&\\
\cline{4-8}
      &&&1.00&1.05&1.10&1.20&1.30&Expt.\\
\hline
$E$&$^6$He&&$-30.32$ & $-30.98$ & $-31.71$ & $-33.45$ & $-35.53$&$-29.27$  \\
$S_{2n}$&$^6$He&&$0.39$  & $1.01$  & $1.72$  & $3.39$  & $5.41$&$0.972$\\ 
\hline
$r_m$&$^6$He&&$2.52$   & $2.41$   & $2.32$   & $2.16$   & $2.05$&\\
    &$^4$He          &&$1.41$   & $1.41$   & $1.41$   & $1.41$   & $1.41$&\\
$r_p$&$^6$He&&$1.90$ & $1.83$& $1.78$& $1.68$& $1.61$&\\
    &$^4$He          &&$1.41$   & $1.41$& $1.41$   & $1.41$   & $1.41$&\\
$r_n$&$^6$He         &&$2.78$   & $2.65$   & $2.54$   & $2.37$   & $2.23$&   \\
    &$^4$He         &&$1.41$   & $1.41$   & $1.41$   & $1.41$   & $1.41$&   \\
$r_{pp}$&$^6$He        &&$2.50$   & $2.49$   & $2.48$   & $2.44$   & $2.39$&   \\
      &$^4$He&&$2.37$   & $2.37$   & $2.37$   & $2.37$   & $2.37$&   \\
\hline\hline
\end{tabular}
\label{phys.tab}
\end{center}
\end{table}

Calculated physical quantities are summarized in Table~\ref{phys.tab}
for different $u$ values.
Our results on binding energies are also consistent with 
those by EIHH~\cite{goerke12} for all $u$ values.
Since the $u$ parameter controls the strength 
of the odd-partial waves, the binding energy and radius 
of $^{4}$He, which has almost $(0s_{1/2})^4$ configuration, 
do not depend on $u$.
Some dependence on $u$ appears for $^{6}$He, however, because
the two valence neutrons move in the $P$ orbit 
according to the simplest shell model. 
A larger $u$ value gives a more attractive $P$-wave interaction.
It is found that the experimental $S_{2n}$ value 
is reproduced with $u=1.05$, so that we use $u=1.05$ in what follows 
unless otherwise mentioned.

Our wave function for $^4$He underestimates its 
point proton rms radius $r_p$ by 0.05 fm~\cite{angeli13}, and partly because of this 
the calculated $r_p$ of $^6$He is slightly smaller than the experimental
values, $1.938(23)$~\cite{brodeur12} and $1.912(18)$~\cite{wang04},
extracted from charge radius measurements. 
The experimental matter radius of $^6$He is not as precise as the proton radius  
and is somewhat scattered: The empirical values are 
2.48(3)~\cite{tanihata88}, 2.30(7)~\cite{alkhazov97}, 
2.33(4)~\cite{tanihata92}, and 2.37(5)~\cite{kiselev05}. The averaged empirical matter radius, 
2.37(10) fm, is consistent with the calculated  matter radius.
Many theoretical works have been devoted to understanding the rms radii of $^6$He 
with different methods and interactions. See, e.g., Ref.~\cite{brodeur12, bacca12} 
and references therein.

The $r_p$ value of $^6$He is not necessarily the same as that 
of $^4$He due to a recoil effect of the core. Even though the $\alpha+n+n$ three-body model is 
quite good for the ground state of $^6$He, the $r_p$ value of $^6$He becomes larger than that of $^4$He 
because the c.m. of the $\alpha$ core moves around the c.m. of $^6$He, and the difference between them 
depends on the extent to which the c.m. of the $\alpha$ core fluctuates.  
Denoting the relative distance vector 
between the core and two valence neutrons by $\bm{R}=\bm{y}_2^{(\rm T)}$, 
in general we obtain the following relation between the $r_p$ values of $A$ and $A-2$ systems~\cite{suzuki91,horiuchi07}
\begin{align}
\left<r_p^2(A)\right>=\left<r_p^2(A-2)\right>+\frac{4}{A^2}\left<\bm{R}^2\right>.
\label{distance.eq}
\end{align}
The second term of the right-hand side of the above equation expresses the recoil effect.
With Eq. (\ref{distance.eq}), 
the expectation value of the relative distance, 
$\sqrt{\left<R^2\right>}$, is estimated as 
3.50\,fm, which is consistent with 3.89\,fm obtained by 
an $\alpha+n+n$ three-body calculation~\cite{horiuchi07}.

The rms distance of the two protons, $r_{pp}$, serves as a measure 
of whether the $^{4}$He core is frozen or not in $^{6}$He. 
If the two-proton configuration of $^{6}$He is the same as that of $^{4}$He,
their $r_{pp}$  values are expected to be the same. 
The $r_{pp}$ of $^6$He is found to be 5\% larger 
than that of $^{4}$He, which indicates a core swelling.   
This effect becomes larger as the valence neutron binding energy decreases.
As pointed out in Ref.~\cite{arai99},  the proton tail of the $\alpha$ core plays a 
vital role in binding the two valence neutrons.
A proton tends to be closer to neutrons rather than to the other proton 
because the $pn$ interaction is stronger than the $pp$ interaction.
In case that the valence neutrons are weakly bound,
the proton in the $\alpha$ core is attracted by the valence
neutrons moving far from the core, and thus $r_{pp}$ of $^{6}$He 
becomes larger than that of the free $\alpha$ particle.
The effect is never taken into account in a macroscopic three-body model
of $\alpha+n+n$ but is realized in a fully microscopic six-body model.

\begin{figure}[ht]
\begin{center}
\epsfig{file=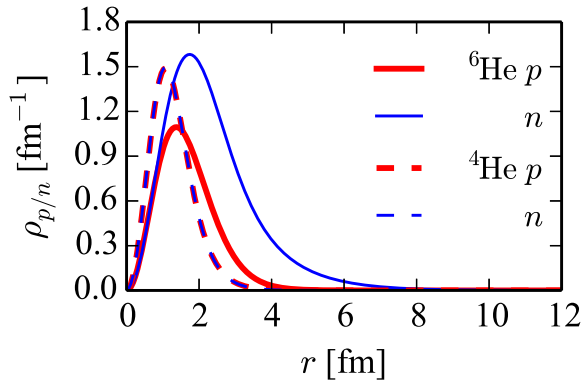,scale=1.1}
\caption{(Color online). Proton and neutron density distributions of $^{4,6}$He.}
\label{dens.fig}
\end{center}
\end{figure}

Figure~\ref{dens.fig} displays the density distributions of $^{4,6}$He defined by
\begin{align}
\rho_{p/n}(r)=
\left<\Psi_{0}\right|\sum_{i\in p/n}\delta(|\bm{r}_i-\bm{x}_{\rm cm}^{(N)}|-r)
\left|\Psi_{0}\right>,
\end{align}
where $\Psi_0$ is the ground state wave function. The density is 
normalized to the number of protons or neutrons.
The density distributions for both proton and neutron
are the same in $^4$He and their peak position is 1.1 fm,
whereas the neutron density of $^6$He, peaked at 1.7 fm, is  
very much extended showing a two-neutron halo feature.
The peak of the proton density of $^6$He is shifted to 1.3 fm
due to the recoil effect,
and the density exhibits more extended distribution than that of $^4$He.

\subsection{Electric dipole strength}
\label{excitation.sec}

\begin{figure*}[th]
\begin{center}
\epsfig{file=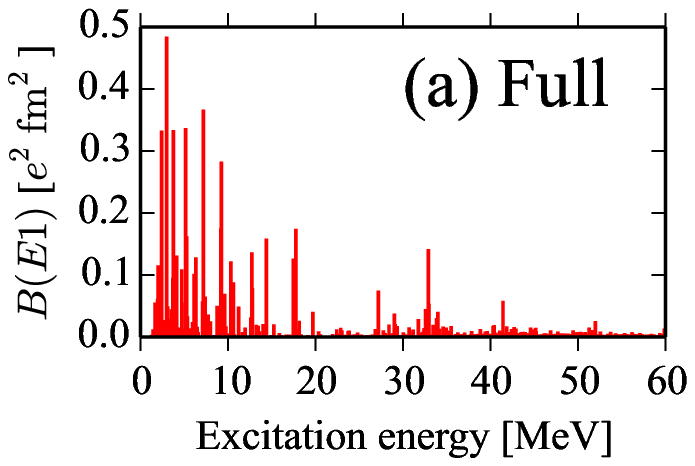,scale=1}
\epsfig{file=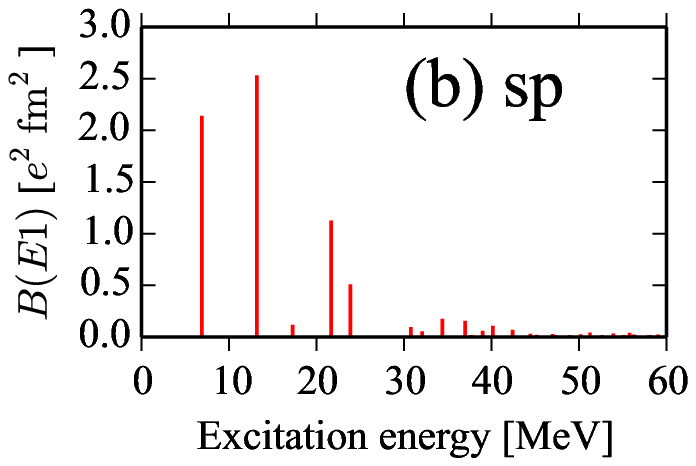,scale=1}
\epsfig{file=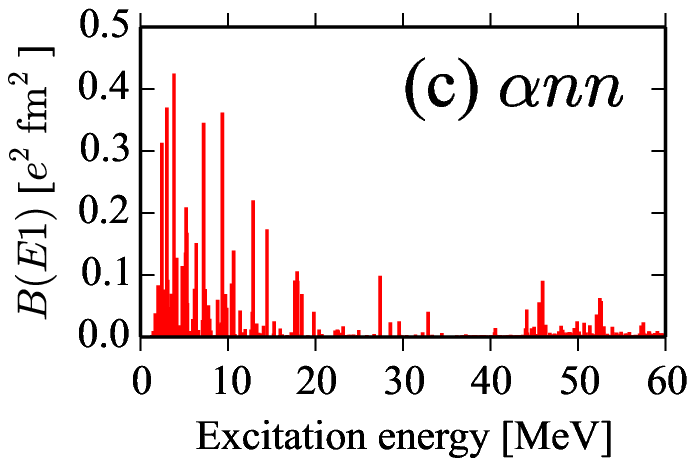,scale=1}
\epsfig{file=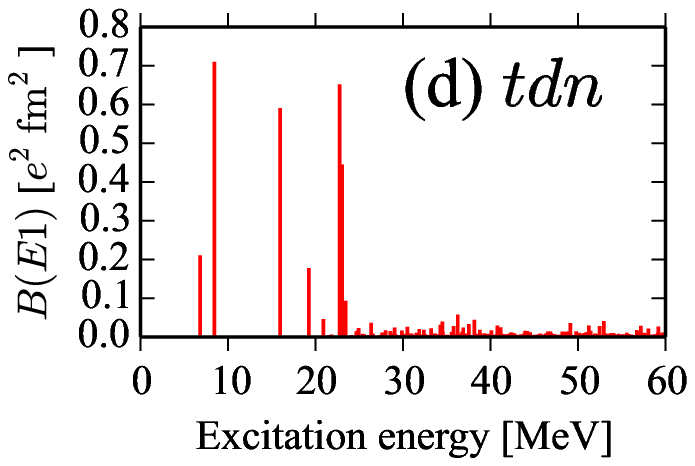,scale=1}
\caption{(Color online). Electric dipole strength of $^6$He calculated with
each basis set of the excitation diagram. 
Full indicates the result of calculation including all the basis sets.
Note the different scale of the sp $B(E1)$ value.}
\label{E1disc.fig}
\end{center}
\end{figure*}

We show  continuum-discretized $E1$ strength.
The Hamiltonian is diagonalized in the basis states that are 
defined in Sec.~\ref{wavefn.sec}, and the reduced transition probability
for the $E1$ operator is calculated by
\begin{align}
B(E1,\nu)=\sum_{M\mu}\left|\right<\Psi_{1M}(E_\nu)|
\mathcal{M}_{1\mu}|\Psi_{0}\left>\right|^2,
\label{be1.eq}
\end{align}
where $\Psi_{1M}(E_\nu)$ is $\nu$th final state wave function with 
excitation energy $E_\nu$.

Figure~\ref{E1disc.fig} displays the $E1$ strength calculated with each model 
space as well as with the Full space.
The sp model space produces two prominent and concentrated peaks
at 6.91 and 13.20 MeV and fragmented strength at the excitation energy of 30-40\, MeV.
The two prominent peaks may correspond to the observed levels
at 5.6 and 14.6 MeV with large decay widths, 12.1 and 7.4 MeV, 
respectively~\cite{tilley02}, though their $J^{\pi}$ value are not identified as $1^-$. 
The fragmented strength at the higher energy region appears to correspond to the GDR.
Compared to the sp result, 
the $\alpha+n+n$ model space presents many much smaller and 
fragmented peaks due to the weak binding nature of the valence neutrons.
Small peaks are concentrated at the low-energy region around 4 MeV.
These peaks in the low-lying $E1$ strength may correspond to the SDM. 
The $t+d+n$ model space presents several concentrated peaks below 24 MeV as
well as many fragmented strength beyond that energy. 
As discussed before, this model space explicitly includes the 
configurations in which two protons can be apart from each other,
 and we will confirm later that this class of model space plays a decisive role in describing  
the GDR due to the core excitations.

We study the mechanism of the appearance of the low-lying $E1$ strength
and note that the SDM discussed here 
does not necessarily mean a narrow resonance.
A microscopic $\alpha+n+n$ cluster model calculation was 
performed in Ref.~\cite{csoto94} to find low-lying $1^-$ resonances
with the complex scaling method, and no such resonant state
appeared in the low-energy region.
As seen in Fig.~\ref{E1disc.fig}(a),
the strength distribution in the low-energy region
is densely obtained but no such peak is found 
that is sufficiently stable against the increase of the model space. 
It is unlikely to obtain a narrow low-lying $1^-$ resonance
consistently with the result of Ref.~\cite{csoto94}.

As noted already, each model space has significant overlap, and in the calculation of 
the Full model space almost all the prominent peaks are fragmented into small strength.
Roughly speaking, the strength distribution  can be divided into
two groups: Broad and strong low-lying strength 
centered around 3 MeV and extending to 20 MeV 
and higher peaks at 33 MeV. It is apparent that the 
$\alpha+n+n$ configurations are of vital importance 
in describing the low-lying $E1$ strength because
the low-lying peaks show up only with the $\alpha+n+n$ configurations. 
The strength at around 33 MeV corresponds to the GDR and apparently appear when 
the core excited configurations are explicitly included.

\begin{figure}[th]
\begin{center}
\epsfig{file=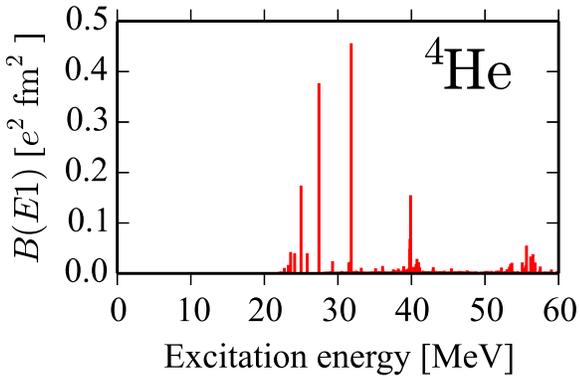,scale=1.1}
\caption{(Color online). Electric dipole strength of $^4$He.}
\label{E1disc4He.fig}
\end{center}
\end{figure}

What is the possible structure of the GDR in $^6$He?
Since the GDR regions already exceed the breakup threshold of $^4$He,
the breaking of the $\alpha$ core should play an essential role.
Figure~\ref{E1disc4He.fig} displays the $E1$ strength 
for $^4$He. The excited $1^-$ states are constructed 
as in Ref.~\cite{horiuchi12a}.
Differently from $^6$He, no strength appears at 
the low-energy region because the threshold of $^3$H$+p$ 
is fairly high. Four prominent peaks are found at 25.0, 27.5, 31.8, and 39.9 MeV,  
and we see that several peaks in $^6$He appear at the positions similar to those energies.
We will confirm in Sec.~\ref{three-body.sec} that the peaks in the region between 
33 and 41\,MeV are due to the excitation of $^4$He. 

The $E1$ sum rule is useful as a qualitative measure of judging the extent to which the model space 
is extensive to describe the $E1$ excitation. If $\Psi_{1M}(E_\nu)$ of Eq.~(\ref{be1.eq}) 
forms a complete set for configurations with $1^-$,  
 a well-known non-energy-weighted sum rule for the $E1$ operator reads 
\begin{align}
\sum_\nu B(E1,\nu)
=e^2\left(Z^2\left<r_p^2\right>-\frac{Z(Z-1)}{2}\left<r_{pp}^2\right>\right),
\label{newsr}
\end{align}
where $Z$ is the number of protons.
The right-hand  side of the above equation depends  only on the ground state properties, 
which turns out to be 7.21 $e^2$fm$^2$. 
We obtain 99.8, 90.4, and 65.8\% of
the sum rule  for the sp, $\alpha+n+n$, and $t+d+n$ configurations, respectively.
The Full space exhausts 99.9\% of the sum rule.
This confirms that our model space is extensive enough to account for 
the configurations excited by the $E1$ operator. It should be noted, however, that the sp 
model space alone accounts for the sum rule but the fragmentation of the $E1$ strength 
is possible only with the coupling to the $\alpha+n+n$ and $t+d+n$ model space. 

\subsection{Comparison with experiment }
\label{photo.sec}

\begin{figure}
\begin{center}
\epsfig{file=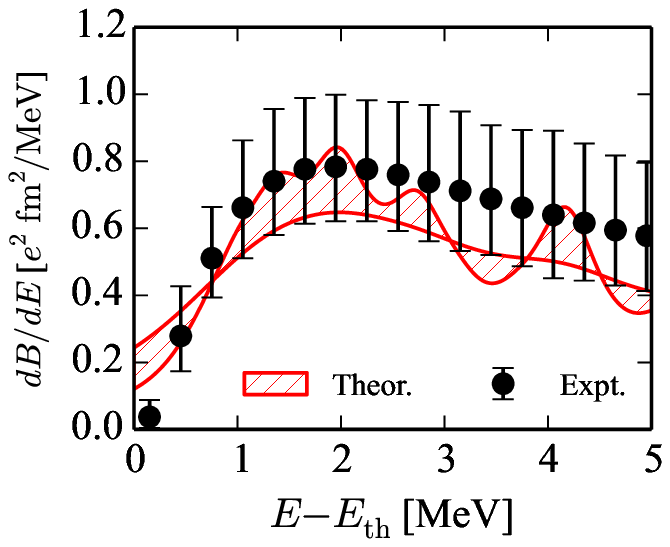,scale=1.15}
\caption{(Color online). 
Electric dipole strength functions of $^{6}$He
measured from the threshold energy $E_{\rm th}$.
A hatched area denotes theoretical uncertainly 
due to a choice of the smearing width ranging from 0.75 to 2.0 MeV
in the Lorentzian function (\ref{lorentz.eq}).
Experimental data are taken from Ref.~\cite{aumann99}.}
\label{dbde1.fig}
\end{center}
\end{figure}

\begin{figure}
\begin{center}
\epsfig{file=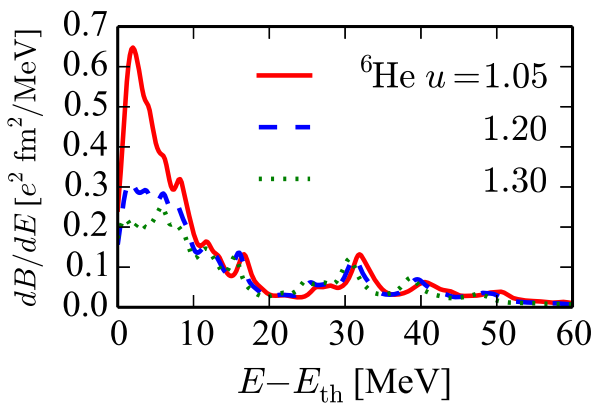,scale=1.15}
\caption{(Color online). The same  as Fig.~\ref{dbde1.fig} 
but for a wider energy region. 
The results with different $u$ parameters of the MN potential
are also plotted. The smearing 
width of the Lorentzian function is set to be 2.0 MeV.
}
\label{dbde2.fig}
\end{center}
\end{figure}

The $E1$ strength function is approximately obtained by a convolution
of $B(E1)$ with the Lorentzian function as follows: 
\begin{align}
\frac{dB(E1,E)}{dE}=\sum_{\nu}N(E_\nu,\Gamma)L(E,E_\nu,\Gamma)B(E1,\nu)
\end{align}
with 
\begin{align}
L(E,E_\nu,\Gamma)=\frac{\Gamma}{2\pi}\frac{1}{(E-E_\nu)^2+(\Gamma/2)^2},\\
N(E_\nu,\Gamma)=\frac{1}{1-\int_{-\infty}^{E_{\rm th}}L(E^\prime,E_\nu,\Gamma)dE^\prime},
\label{lorentz.eq}
\end{align}
where $N(E_\nu,\Gamma)$ is introduced to renormalize the strength near the two-neutron 
threshold of $^6$He, $E_{\rm th}$, satisfying  
$\int_{E_{\rm th}}^\infty  N(E_\nu,\Gamma)L(E,E_\nu,\Gamma)dE=1$. 
Figure~\ref{dbde1.fig} compares the calculated $E1$ strength function
with that  extracted from the Coulomb breakup measurement~\cite{aumann99}.
The hatched band indicates the variation of theoretical strength functions for different width parameters, $\Gamma$=0.75 to 2.0 MeV.
The range of $\Gamma$ is chosen referring to Ref.~\cite{pinilla11}.
An oscillatory behavior of the $E1$ strength is observed with 
the small $\Gamma$ but the $\Gamma$ dependence becomes small 
for $\Gamma> 1$ MeV. In view of large error bars of the data we conclude 
that the calculated $E1$ strength 
function fairly well agrees with experiment.

Figure~\ref{dbde2.fig} displays the $E1$ strength function in wider energy range.
The width $\Gamma$ is taken as 2.0 MeV.
The peak at the lower energy is expected to have the SDM structure while 
the other has the GDR structure as implied by Ref.~\cite{bacca02}.
We examine the sensitivity of the $E1$ strength function to the $u$ parameter. 
The  peak position at the low excitation energy shifts drastically to higher energy and its strength becomes  much smaller with increasing $u$. This  
is easily understood because the valence neutrons get more binding and are 
not easily excited. See Table~\ref{phys.tab}. No drastic change of the peak position and strength 
is found in the GDR region around 33 MeV where
the $\alpha$ breaking is dominant. As confirmed above, the large enhancement of the $E1$ strength 
at low energy is due to the weakly bound neutrons, and therefore to reproduce 
the experimental two-neutron separation energy is essential to account for the low-lying $E1$ strength.
We obtain some other peaks at 10 and 18 MeV as seen in 
Fig.~\ref{dbde2.fig}.  We will discuss their structure in Sec.~\ref{softmode.sec}.

\subsection{Soft and giant dipole excitation modes}
\label{softmode.sec}

\begin{figure}[ht]
\begin{center}
\epsfig{file=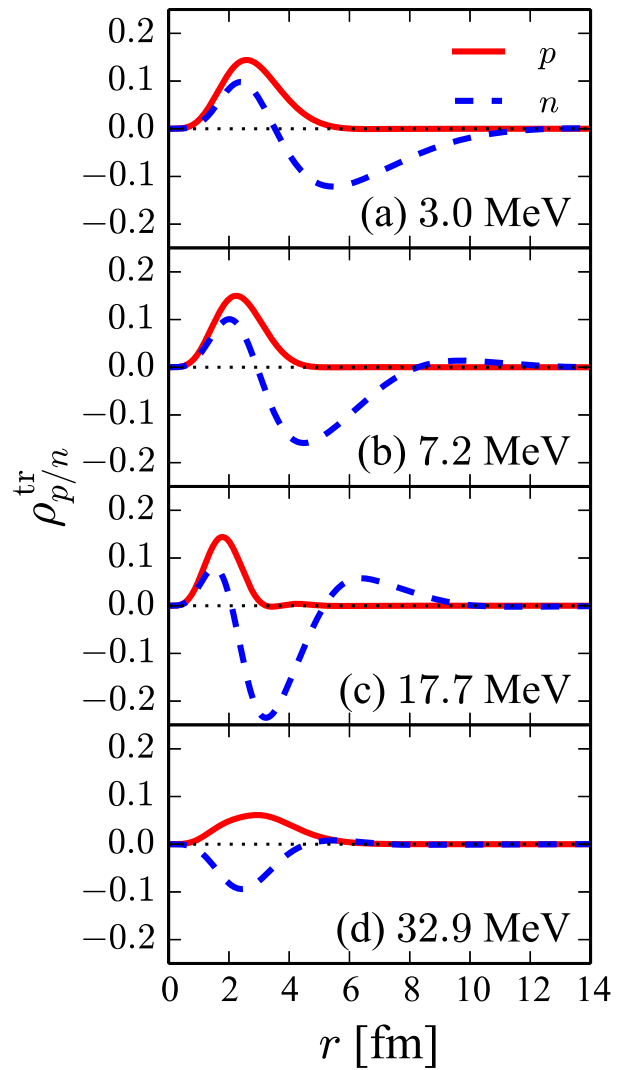,scale=1.15}
\caption{(Color online). Electric dipole transition densities
for proton and neutron of the states with the excitation energy of
(a) 3.0, (b) 7.2, (c) 17.7, and (d) 32.9 MeV, respectively.}
\label{tdens.fig}
\end{center}
\end{figure}

To discuss how the protons and neutrons respond in the $E1$ excitation, we calculate 
the $E1$ transition density 
\begin{align}
&\rho^{\rm tr}_{p/n}(E_\nu,r)\notag\\
&=
\left<\Psi_{1}(E_\nu)\right\|\sum_{i\in p/n}\mathcal{Y}_1(\bm{r}_i-\bm{x}_6)\delta(|\bm{r}_i-\bm{x}_6|-r)\left\|\Psi_{0}\right>.
\end{align}
As defined in Eq.~(\ref{E1.eq}), the $E1$ transition matrix element is given by  
the proton transition density as 
\begin{align}
\left<\Psi_{1}(E_{\nu})\left\|\mathcal{M}_{1}\right\|\Psi_0\right>=
e\sqrt\frac{4\pi}{3}\int_0^\infty \rho^{\rm tr}_{p}(E_\nu,r)dr.
\label{trdene1.eq}
\end{align}
If the $\alpha+n+n$ three-body model is assumed and no excitation of the $\alpha$ core 
is allowed, the neutron transition density is related to the proton transition density by
\begin{align}
\rho^{\rm tr}_{n}(E_\nu,r)=\rho^{\rm tr}_{p}(E_\nu,r)+\rho^{\rm tr}_{2n}(E_\nu,r),
\label{np.tr.density}
\end{align}
where $\rho^{\rm tr}_{2n}(E_\nu,r)$ stands for a contribution from the halo neutrons and it is 
entirely determined by the two-neutron wave function.

Figure~\ref{tdens.fig} plots the proton and neutron transition densities
of the most prominent peaks at about 3, 7, 18, and 33\,MeV. 
In the lowest prominent peak at 3.0 MeV, 
the transition densities of proton and neutron coincide 
up to about 1.7 fm, which corresponds to 
the peak of the ground state neutron density (See Fig.~\ref{dens.fig}). 
The neutron transition density deviates from the proton one beyond 1.7 fm,
showing in-phase oscillation in the interior region, $r \lesssim 3$ fm, 
and out-of-phase oscillation in the external region $r \gtrsim 3$ fm.  
Furthermore, the neutron transition density is very much extended
beyond 10 fm. These observations agree with what we expect 
from the relationship~(\ref{np.tr.density}) for an ideal SDM and 
are consistent with the classical interpretation of the SDM discussed 
in Ref.~\cite{suzuki90a}.
As the energy increases up to 18 MeV, the 
proton transition density gradually shrinks but its basic pattern is still 
kept. This shrinkage is due to the fact that the $\rho^{\rm tr}_{2n}(E_\nu,r)$ 
of the neutron transition density vibrates 
more and more rapidly with an increase of the number of nodes of oscillation. Though the 
penetration of the valence neutrons into the internal region grows gradually, 
the internal structure characteristic of the SDM 
does not change so much as shown in Figs.~\ref{tdens.fig} (b) and (c).
The oscillatory behavior becomes even stronger at 17.7 MeV, 
showing little distortion of the core. 
The neutron transition density reaches a maximum at 3.4 fm. The protons 
follow the motion of the neutrons, and 
the resulting proton transition density exhibits a destructive pattern.
Such neutron oscillation leads to the smaller $E1$ strength  
than that of the lowest-lying peak.

\begin{figure}[ht]
\begin{center}
\epsfig{file=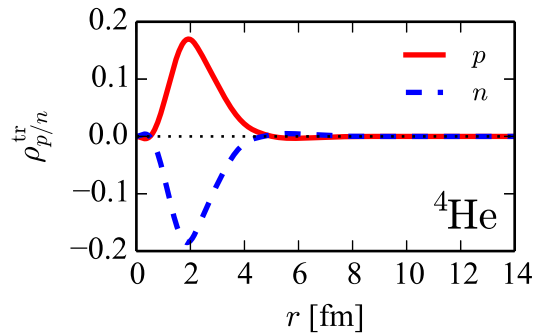,scale=1}
\caption{(Color online). Electric dipole transition densities
for proton and neutron of $^4$He 
at the excitation energy of 31.8 MeV.}
\label{tdens4He.fig}
\end{center}
\end{figure}

At the higher peak with 33\,MeV, the transition densities
clearly show the out-of-phase oscillation in the whole region,
which is typical of the GDR. 
The proton transition density shows somewhat broader distribution than those
of the low-lying states. This suggests the strong distortion of the core 
due to the dipole field, which can never be described by 
$\alpha+n+n$ three-body models with an inert core.
As an example of the ideal GDR, we plot in 
Fig.~\ref{tdens4He.fig} the transition densities of 
the most prominent strength of $^4$He at 31.8 MeV. See Fig.~\ref{E1disc4He.fig}.
Both the proton and neutron transition densities 
show the identical distribution with opposite phases.
They are peaked at 1.9 fm, which is further inside than those of the GDR of $^6$He.

We also investigate the excited modes of $^6$He at about 40 MeV and find that 
their transition densities are similar to that of the GDR but have  
more oscillations and smaller amplitudes.

\subsection{Validity of $\alpha+n+n$ three-body picture}
\label{three-body.sec}

To examine how good the three-body model of $\alpha+n+n$ is for $^{6}$He, 
we study the cluster sum rule~\cite{alhassid82} as a qualitative measure. 
We rewrite the $E1$ operator (\ref{E1.eq}) as 
\begin{align}
\mathcal{M}_{1\mu}=e\left(
\sum_{i\in p}(\bm{r}_i-\bm{x}_{\rm cm}^{(A-2)})_\mu-\frac{2Z}{A}\bm{R}_\mu\right).
\end{align}
If the system has a core plus
two-nucleon structure, the first term on the right-hand side of the above equation does not
contribute to the $E1$ strength, which leads to the non-energy-weighted cluster sum rule~\cite{sagawa90,suzuki90b}
\begin{align}
 B(E1;{\rm NEWCSR})=e^2\left(\frac{2Z}{A}\right)^2\left<\bm{R}^2\right>.
\label{clusum.eq}
\end{align}
We calculate a cumulative sum of the $E1$ strength, 
$\sum_{\nu=1}^{\nu_{\rm max}}B(E1,\nu)$, and find that the cumulative sum 
exceeds the $B(E1;{\rm NEWCSR})$ value 
of 5.44\,$e^2$fm$^2$ at $E_{\nu_{\rm max}}=26.8$\,MeV.  
Therefore the cluster sum rule occupies about 75\% of the non-energy-weighted sum rule~(\ref{newsr}) and its strength appears below the excitation energy of 25\,MeV. Since the GDR appears 
above 30\,MeV, we conclude 
that the SDM and GDR  are well separated
and the low-lying strength is understood with the $\alpha+n+n$ 
three-body structure.

\begin{figure}[th]
\begin{center}
\epsfig{file=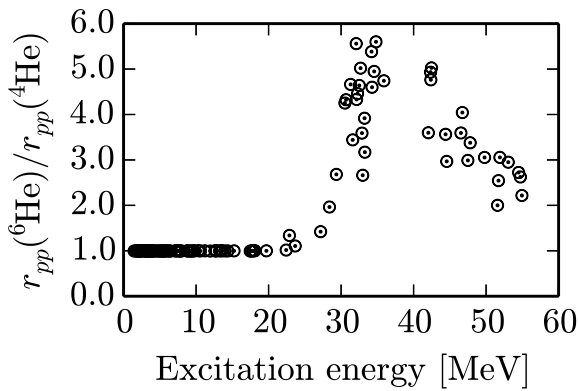,scale=1.1}
\caption{Ratio of the proton-proton ($pp$) distance of $^6$He
to that of the ground state of $^4$He as a function of $E1$ excitation energy.}
\label{pp.fig}
\end{center}
\end{figure}

The above conclusion is further confirmed by calculating the proton-proton rms  distance, $r_{pp}$, 
of the  state $\Psi_{1M}(E_\nu)$  
because $r_{pp}$ can be a measure of whether or not the $\alpha$ core exists in $^6$He.
By selecting those states that have 1/1000 of the $E1$ sum rule value, we plot in  
Fig.~\ref{pp.fig}  the ratio 
of $r_{pp}$ of $^6$He to that of $^{4}$He.
The ratio is unity up to the excitation energy of about 20 MeV that corresponds to
the excitation energy of the first excited $0^+$ state of $^{4}$He 
(20.21 MeV~\cite{tilley92}).  This indicates that the core is not virtually excited, and thus  
the three-body picture holds very well in the low-energy region below 20 MeV. 
As discussed in Sec.~\ref{ground.sec}, the ground state of $^6$He has  
5\% larger $r_{pp}$ than that of $^4$He. 
No such core swelling effect is found, however, in the low-lying $1^-$ states. 
In the $1^-$ states the valence neutrons are further away from the core 
and receive essentially no interaction from the core.
Beyond the excitation energy of 20 MeV, the ratio suddenly increases  
and reaches a maximum at the GDR region, indicating a large distortion of the core.  

In Ref.~\cite{bacca02} the two peak structure of
the $E1$ strength function of $^{6}$He is discussed from the isospin 
decomposition to $T=1$ and 2 states. 
The low-lying peak is dominated by the total isospin $T=1$ state, and
above $\sim$20 MeV the $T=2$ component contributes as well. 
Since the three-body structure of $\alpha+n+n$ consists of $T=1$, our finding in the $r_{pp}$ analysis 
is consistent with the result of Ref.~\cite{bacca02}.

The core excitation often plays an important role in enhancing 
the low-lying $E1$ strength of two-neutron halo nuclei.
For example, in the case of $^{11}$Li,
core excitations produce a large admixture of the 
$(1s_{1/2})^2$ component in the ground state~\cite{thompson94,varga02,myo07}. 
Such very extended $S$-orbitals enhance the matter radius
as well as the low-lying $E1$ strength~\cite{myo07,kikuchi12}.
In Ref.~\cite{inakura14}, the $E1$ strength function of $^{22}$C
is discussed with the Skyrme-Hartree-Fock method 
on a 3-dimensional coordinate space and the core ($^{20}$C) excitation is found to play 
an important role to account for the $E1$ strength. 
The transition densities for the PDR and GDR regions 
shown in the paper are similar to those obtained 
in the present work. There is however a clear difference between $^{22}$C and $^6$He.  
In the case of $^{22}$C the single-particle energies of the $0d_{5/2}$ and $1s_{1/2}$ orbitals are very close,
and the nucleons in the $0d_{5/2}$ orbits 
can easily be excited to continuum by the $E1$ operator, which is a dominant process of the core 
excitation in the low-energy region. In 
the case of $^{6}$He, however, the $\alpha$ core is not easily excited and 
the energy gap between the $0s_{1/2}$ and $0p_{3/2}$ orbitals is fairly large. As a result 
of both effects the motion of the core and valence neutrons decouples approximately 
in the low-lying $E1$ excitation.

The large enhancement of the low-lying $E1$ transition in
the two-neutron halo nuclei is mainly due to the weak binding feature of 
the valence nucleons, but a special care 
about the structure of the core is needed for understanding its origin.

\subsection{Compressional $E1$ mode}
\label{comp.E1}

We examine another $E1$ mode, the so-called compressional $E1$ ($cE1$) mode~\cite{stringari82}. 
The operator for the mode is defined as
\begin{align}
\mathcal{M}^{\rm comp.}_{1\mu}=e\sum_{i\in p}|
\bm{r}_i-\bm{x}_6|^3{Y}_{1\mu}(\widehat{\bm{r}_i-\bm{x}_6}).
\end{align}
According to a simple harmonic-oscillator shell model,
the $cE1$ mode appears at the high-energy region
because it requires at least  3$\hbar\omega$ 
excitations 
from the $0\hbar\omega$ ground state. 
However, if the ground state contains some amount of correlated components, 
the mode may appear at the low-energy region 
by coupling with higher oscillator shells 
that are already incorporated in the ground state wave function. The matrix element of 
$\mathcal{M}^{\rm comp.}_{1\mu}$ is given by the proton transition density as
\begin{align}
\left<\Psi_{1}(E_{\nu})\left\|\mathcal{M}^{\rm comp.}_{1}\right\|\Psi_0\right>=
e\int_0^\infty r^2 \rho^{\rm tr}_{p}(E_\nu,r)dr.
\end{align}
The matrix element for the isoscalar (IS) compressional dipole 
($c1$) mode is calculated as 
\begin{align}
\int_0^\infty r^2 
\Big(\rho^{\rm tr}_{p}(E_\nu,r)+\rho^{\rm tr}_{n}(E_\nu,r)\Big)dr.
\label{isoscalarcomp}
\end{align}
Since the transition 
is not necessarily induced by the electromagnetic
interaction but the nuclear one, we omit $e$ from the matrix element.

\begin{figure}[th]
\begin{center}
\epsfig{file=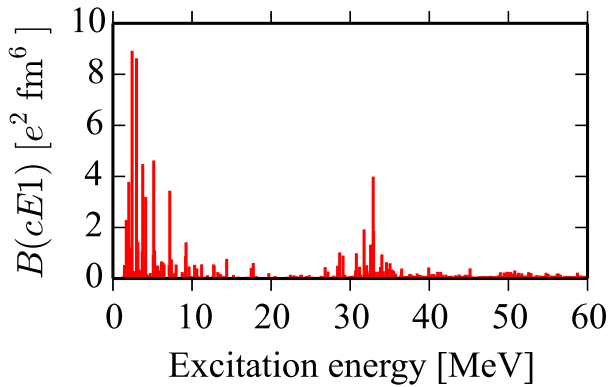,scale=1.15}
\caption{(Color online). Compressional electric dipole strength 
of $^{6}$He as a function of the excitation energy.}
\label{compe1.fig}
\end{center}
\end{figure}

\begin{figure}[th]
\begin{center}
\epsfig{file=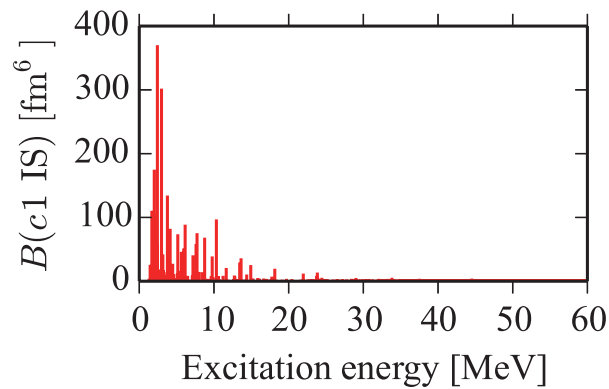,scale=1.15}
\caption{(Color online). Isoscalar compressional dipole strength 
of $^{6}$He as a function of the excitation energy.}
\label{compe1is.fig}
\end{center}
\end{figure}

Figure~\ref{compe1.fig} displays the $cE1$ strengths
as a function of the excitation energy.
Similarly to the normal $E1$ operator, two peak structure is found
but the low-lying peaks are more concentrated and enhanced. 
The SDM is more characterized by the $cE1$ mode
rather than the $E1$ one by the additional $r^2$ factor of the operator.
As Eq.~(\ref{isoscalarcomp}) suggests, 
the IS $c1$ mode disappears at the 
GDR region because of the cancellation 
of the proton and neutron transition densities.
As shown in Fig.~\ref{compe1is.fig},
the IS $c1$ strength shows up only in the low-energy region
and has a strong peak at 2.4 MeV. A measurement of
the IS $c1$ mode is of particular interest in relation to the SDM.
One possible way to excite the IS $c1$ mode is inelastic $\alpha$
scatterings, $^6$He$(\alpha,\alpha^\prime)^6$He($1^-$).

\section{Conclusions}

\label{conclusions.sec}

We have performed a fully microscopic six-body calculation 
to explore the electric dipole excitation mode in $^6$He.
The ground state wave function is expressed with
the explicitly correlated Gaussians.
The model space responsible for the $E1$ excitation is 
also expressed as a combination of the correlated Gaussians 
with the global vector.
The model space explicitly incorporates the configurations for describing 
the single-particle excitation as well as the final state correlations 
of $\alpha+n+n$ and $t+d+n$ decay channels.

The ground state properties of the two-neutron separation energy and 
matter and proton radii and the low-lying $E1$ strength are 
all reproduced consistently with the observations. 
The ground state structure is well understood with 
the $\alpha+n+n$ three-body model though 
a few-percent core swelling is produced by the halo neutrons. 

It is found that the $E1$ non-energy-weighted 
sum rule is fully accounted for by our model space.
The $E1$ strength function 
exhibits two-peak structure at around 3 and 33 MeV excitation energy.
The lower peak is well understood in the framework of the $\alpha+n+n$ structure
and its excitation mechanism is consistent with the classical interpretation 
of the soft dipole mode (SDM), 
in which in-phase proton-neutron oscillation occurs 
in the internal region  
whereas out-of-phase oscillation occurs in the surface region. Beyond the surface 
region the neutron transition density extends to large distances.  
The higher peak is the typical giant dipole resonance that 
exhibits out-of-phase proton-neutron oscillation in the whole region. 
Just a few MeV above the SDM peak, we find some new modes that can be regarded as 
a vibrational excitation of the SDM. 
A measurement for such mode is interesting. 
We find out that the SDM may be more apparently disclosed
by the isoscalar compressional dipole transition rather than
the $E1$ transition, and point out the possibility of observing it
by inelastic $\alpha$ scatterings.

In this study, we have succeeded to describe, in a single scheme, the $E1$ 
excitation in a wide energy region from pigmy to giant dipole resonance. 
It is interesting to extend it to other multipoles or systems
to explore other new modes.

\section*{Acknowledgments}

The work was in part supported by JSPS KAKENHI Grant Numbers (24540261 and
25800121)

\end{document}